# Quadratic Band Touching and Nontrivial Winding Reveal Generalized Angular Momentum Conservation

Yihan Wang[1†], Domenico Bongiovanni[1,2†], Dario Jukić[3], Sihong Lei[1], Zhichan Hu[1], Daohong Song[1,4*], Jingjun Xu[1], Roberto Morandotti[2], Hrvoje Buljan[1,5], and Zhigang Chen[1,4*]

[1]*The MOE Key Laboratory of Weak-Light Nonlinear Photonics, TEDA Applied Physics Institute and School of Physics, Nankai University, Tianjin 300457, China*

[2]*INRS-EMT, 1650 Blvd. Lionel-Boulet, Varennes, Quebec J3X 1S2, Canada*

[3]*Faculty of Civil Engineering, University of Zagreb, A. Kačića Miošića 26, Zagreb 10000, Croatia*

[4]*Collaborative Innovation Center of Extreme Optics, Shanxi University, Taiyuan, Shanxi 030006, China*

[5]*Department of Physics, Faculty of Science, University of Zagreb, Bijenička c. 32, Zagreb 10000, Croatia*

[†]*These authors contributed equally to this work*

*\*e-mail: songdaohong@nankai.edu.cn, zgchen@nankai.edu.cn*

**Angular momentum conservation stands as one of the most fundamental and robust laws of physics. In discrete lattices, however, its realization can deviate markedly from the continuous case, especially in the presence of nontrivial momentum-space band touchings. Here, we investigate angular momentum conservation associated with quadratic band-touching points (QBTPs) in two-dimensional lattices. We show that, unlike in graphene lattices hosting linear band-touching points (LBTPs), the conventional angular momentum is no longer conserved near QBTPs. Instead, we identify a generalized total angular momentum (GTAM) that remains conserved for both LBTPs and QBTPs, inherently determined by the topological winding number at the band-touching point (BTP). Using a photonic Kagome lattice, we experimentally demonstrate GTAM conservation through pseudospin–orbital angular momentum conversion. Furthermore, we show that this conservation principle extends to a broad class of discrete lattices with arbitrary pseudospin textures and higher-order winding numbers. These results reveal a fundamental link between pseudospin, angular momentum, and topology, establishing a unified framework for angular-momentum dynamics in discrete systems.**

**Keywords:** Generalized total angular momentum, Quadratic band-touching point, Dirac point, Pseudospin, Winding number, Photonic Kagome lattice, Optical vortices

Crystalline lattices hosting band-touching points (BTPs) provide fertile ground for exploring exotic phenomena governed by Dirac equations [1,2]. A prototypical example is graphene, whose band structure features Dirac points where two linearly dispersive bands touch in momentum space. Electrons excited near these *linear band-touching points* (LBTPs) behave as massless Dirac fermions [1,3,4], giving rise to a pseudospin degree of freedom that mimics real spin in the Dirac equation [5]. The presence of LBTPs underlies a variety of remarkable relativistic and topological phenomena, such as Klein tunneling [6,7] and the half-integer quantum Hall effect [8,9]. These discoveries have inspired intensive studies of Dirac-like relativistic quasiparticles and their topological phases across diverse artificial lattice systems - from electronic and atomic lattices to photonic platforms [10].

Beyond LBTPs, crystalline lattices often host other types of singular BTPs that are not captured by the Dirac framework, giving rise to unconventional quasiparticle analogs with no counterparts in high-energy physics [11]. Such unconventional band degeneracies have spurred wide exploration in both two- and three-dimensional systems [2]. In two dimensions, for example, one notable case is the *quadratic band-touching point* (QBTP), where the energy gap between two bands varies quadratically with quasi-momentum. QBTPs arise in a variety of symmetry-protected lattices, including Kagome lattices [12–14], $p$-band honeycomb lattices [15,16], bilayer graphene [17–19], and graphene under non-Abelian gauge fields [20]. These quadratic degeneracies lead to unique physical responses, such as anti-Klein tunneling [21], unconventional quantum Hall effects [12,22], and anomalous Landau-level quantization [23], all distinct from their counterparts in graphene lattices with LBTPs. Experimental realizations of QBTPs in cold atoms [16] and photonic systems [24–27] now provide effective platforms for exploring physics beyond Dirac cones.

Since QBTPs cannot be described by Dirac equations, it remains essential to uncover the physical laws that govern their dynamics. In the nonrelativistic limit of the Dirac equation, the inclusion of the Thomas term introduces spin–orbit coupling, leading to the conservation of total angular momentum (TAM), composed of spin angular momentum (SAM) and orbital angular momentum (OAM). An analogous conservation law holds for LBTPs in graphene lattices, where the pseudospin angular momentum (PSAM) effectively acts as SAM [5], as demonstrated for both pseudospin-1/2 [28–31] and higher-pseudospin Dirac-like systems with LBTPs [29,32–35]. However, for excitations near QBTPs with nonlinear dispersion, whether TAM can remain conserved is an open question. In principle, exact conservation laws should be derived from discrete versions of Noether's theorem, yet defining a single, well-behaved conserved quantity that plays the role of continuous angular momentum under discrete lattice symmetries is highly nontrivial. This raises fundamental questions: Can angular momentum be defined in the same sense as for LBTPs in graphene? If conventional angular momentum conservation breaks down, what is the appropriate conserved quantity for QBTPs possessing rotational symmetry? More broadly, can a generalized form of angular momentum conservation exist for higher-order nonlinear BTPs beyond the Dirac paradigm?

In this Letter, we investigate angular momentum conservation in two-dimensional (2D) lattices hosting nonlinear BTPs. We show that the conventional TAM defined within the Dirac framework fails to remain conserved in QBTP lattices. To resolve this, we introduce a *generalized total angular momentum* (GTAM), $\boldsymbol{J}^{(w)} = \boldsymbol{L} + w\boldsymbol{S}$, which incorporates the topological winding number $w$ associated with a BTP. This GTAM consists of pseudospin angular momentum $\boldsymbol{S}$ and orbital angular momentum $\boldsymbol{L}$ that remains conserved for QBTPs, while naturally reducing to the familiar angular momentum in the case of LBTPs. Experimentally, we employ a photonic Kagome lattice that simultaneously hosts LBTPs and QBTPs, which correspond to pseudospin-1/2 states with distinct topological winding numbers $w = 1$ and $w = 2$, respectively. Using such a photonic platform, we directly observe pseudospin-to-OAM conversion and verify the GTAM conservation law. Furthermore, we extend this concept to discrete lattices with arbitrary pseudospin textures and higher winding numbers, establishing a unified framework for angular-momentum conservation in lattices with complex BTPs.

We start the analysis by considering the conservation of TAM in 2D discrete lattices hosting different BTPs described by a pseudospin-1/2 system. The excitation dynamics around BTPs can be captured by the general effective Hamiltonian, given by [29]

$$H_{\text{eff}}^{(m)}(\boldsymbol{p}) = \kappa \begin{pmatrix} 0 & (p_x - ip_y)^m \\ (p_x + ip_y)^m & 0 \end{pmatrix} = \kappa p^m \left( \sigma_x \cos(m\theta) + \sigma_y \sin(m\theta) \right), \quad (1)$$

where $\kappa$ is a coefficient depending on the lattice geometry, $m$ is a positive integer representing the order of the band dispersion, $\theta = \arctan(p_y/p_x)$, and $\boldsymbol{p} = (p_x, p_y)$ is the quasi-transverse momentum displacement vector away from a BTP. The eigenvalues of the effective Hamiltonian $H_{\text{eff}}^{(m)}(\boldsymbol{p})$ are $E_\pm = \pm \kappa p^m$, revealing that the band structures around the BTP exhibits an $m$-order dispersion with quasi-momentum $\boldsymbol{p}$. When $m = 1$, the effective Hamiltonian turns into a Dirac-like Hamiltonian describing the LBTP. If $m > 1$, it describes BTPs formed by nonlinear dispersive bands beyond the Dirac equation. $\sigma_x$ and $\sigma_y$ are the $x, y$ components of the Pauli matrix $\boldsymbol{\sigma}$. For any two-fold BTP (i.e., band touching involves two bands), the PSAM operator is defined as $\boldsymbol{S} = \boldsymbol{\sigma}/2$, where $S_x = \sigma_x/2$, $S_y = \sigma_y/2$, and $S_z = \sigma_z/2$ represent the $x, y$, and $z$ components of $\mathbf{S}$, respectively. The pseudospin eigenstates $\chi_{1/2,\pm 1/2}$ of $S_z$ can be calculated from $\mathbf{S}^2\chi_{1/2,\pm 1/2} = S(S+1)\chi_{1/2,\pm 1/2}$, and $S_z\chi_{1/2,\pm 1/2} = s\chi_{1/2,\pm 1/2}$, where $S$ and $s$ are the eigenvalues of the total and z-component of PSAM, respectively. In the framework of the Dirac equation, it is known that the TAM operator is $\mathbf{J} = \mathbf{L} + \mathbf{S}$, where $\mathbf{L} = \mathbf{r} \times \mathbf{p}$ denotes the OAM operator. For LBTP ($m = 1$) with rotational symmetry in $p_x$-$p_y$ plane, the corresponding z-component $J_z$ remains invariant, given by the commutation relation $\left[H_{\text{eff}}^{(1)}, J_z\right] = 0$, as discussed in [5]. However, the TAM operator $\mathbf{J} = \mathbf{L} + \mathbf{S}$ is not a conserved quantity for nonlinear BTPs because they are not characterized by the Dirac equation (i.e. $\left[H_{\text{eff}}^{(m)}, J_z\right] \neq 0$, $m > 1$), which implies that the conventional TAM conservation law is invalid for nonlinear BTPs. Nevertheless, since the Hamiltonian of the nonlinear BTPs still preserves rotational symmetry, there should be a conserved quantity analogous to angular momentum. Thus, we introduce a new vector quantity, defined as the GTAM expressed as

$$\mathbf{J}^{(w)} = \mathbf{L} + w\mathbf{S}. \tag{2}$$

The GTAM operator satisfies all angular momentum commutation relations as analytically demonstrated in Supplementary Information (SI), confirming its angular momentum nature. Compared to conventionally defined TAM, the GTAM incorporates a coefficient $w$ - the winding number of the BTP, which captures its topological nature and is given by [36]

$$w = \frac{1}{2\pi}\int_0^{2\pi}\left(\widetilde{\mathbf{B}^m} \times \frac{d}{d\theta}\widetilde{\mathbf{B}^m}(\theta)\right)d\theta = m, \tag{3}$$

where $\widetilde{\mathbf{B}^m}(\mathbf{p}) = \mathbf{B}^m(\mathbf{p})/|\mathbf{B}^m(\mathbf{p})|$, and $\mathbf{B}(\mathbf{p}) = (p^m\cos(m\theta), p^m\sin(m\theta), 0)$ represents the pseudomagnetic-field around the BTP that resides in the transverse pseudospin plane. For any $m$, the commutation relation between $H_{\text{eff}}^{(m)}$ and z-component of GTAM $J_z^{(w)}$ is now satisfied: $\left[H_{\text{eff}}^{(m)}, J_z^{(w)}\right] = 0$, indicating that the z-component of GTAM always remains conserved (see proof in SI). Clearly, the angular momentum defined for LBTP can be considered as a special case of GTAM when $m = 1, w = 1$. We emphasize that the rotational symmetry of the BTP protects GTAM conservation, which jointly involves the OAM, PSAM, and the topological winding number for both linear and nonlinear BTPs.

Next, we employ a Kagome lattice featuring both pseudospin-1/2 LBTP and QBTP as a representative model to verify the conservation of GTAM. The Kagome lattice is formed by corner-sharing triangles, as illustrated in Fig. 2(a1). Each unit cell hosts three sites, labeled as A, B, and C, with $t$ being the hopping amplitude between nearest-neighbor sites. $\mathbf{a_1} = 2d\hat{x}$,

$a_2 = -dx + \sqrt{3}dy$, and $a_3 = -dx - \sqrt{3}dy$ are the primitive vectors, where $d$ is the lattice spacing. The analytical form of the tight-binding Bloch Hamiltonian of the Kagome lattice is reported in SI.

The band structure of the Kagome lattice supports two dispersive bands plus one flat band at the bottom [Fig. 2(b1, b2)]. It exhibits six LBTPs at the inequivalent $K'$- and $K$-points, where two uppermost dispersive bands meet linearly at corners of the first Brillouin zone (BZ), and a QBTP at the $\Gamma$-point [Fig.2(b2)] due to band touching between the middle dispersive band and the bottom flat band. Both LBTPs and QBTPs in Kagome lattice can be effectively described by a pseudospin-1/2 Hamiltonian. Thus, the Kagome lattice can serve as an effective model to simultaneously explore the conservation of GTAM for both LBTPs and QBTPs.

Specifically, the excitation dynamics around the LBTPs at $K$-point is governed by the effective pseudospin-1/2 Hamiltonian

$$H_{\text{eff}}^K(\boldsymbol{p}) = \begin{pmatrix} t & -\sqrt{3}tP_- \\ -\sqrt{3}tP_+ & t \end{pmatrix}, \qquad (4)$$

where $P_\pm = p_x \pm ip_y$ (see SI for details). $H_{\text{eff}}^K$ only exhibits the first-order term of momentum $\boldsymbol{p}$, corresponding to the special case of Eq. (1) with $m = 1$. The pseudomagnetic field completes one rotation around the LBTP [Fig. 2(d1)], corresponding to the winding number $w^K = 1$. Therefore, the GTAM operator associated with the LBTP is $\boldsymbol{J}^{(1)} = \boldsymbol{L} + \boldsymbol{S}$ [Fig. 2(c1)]. For LBTP, the pseudospin eigenstates of $S_z$ are $|\psi_1^K\rangle = (1\ 1\ 1)^T$ and $|\psi_2^K\rangle = (e^{i\pi 4/3}\ e^{i\pi 2/3}\ 1)^T$ on the sublattice base, and they are extended over all three sublattices of the Kagome lattice. The conservation of GTAM can be readily verified by obtaining $\left[H_{\text{eff}}^K, J_z^{(1)}\right] = 0$, proving that $J_z^{(1)} = L_z + S_z$ is indeed a conserved quantity for the LBTP.

For excitation around the QBTP at $\Gamma$-point, the dynamics is instead described by the effective pseudospin-1/2 Hamiltonian

$$H_{\text{eff}}^\Gamma(\boldsymbol{p}) = \begin{pmatrix} -2t + tp^2 & -tP_-^2 \\ -tP_+^2 & -2t + tp^2 \end{pmatrix}. \qquad (5)$$

$H_{\text{eff}}^\Gamma(\boldsymbol{p})$ only contains the second-order terms of the momentum $\boldsymbol{p}$, following the form in Eq. (1) with $m = 2$ (see details in SI). Note that $H_{\text{eff}}^\Gamma(\boldsymbol{p})$ is not strictly chiral due to the diagonal term $tp^2$ arising from the flat band [37]. However, this term does not affect the rotation symmetry of the Hamiltonian and thus leaves the GTAM conservation intact. Different from the LBTP case, the pseudomagnetic-field of the QBTP completes a double rotation, resulting in a winding number $w^\Gamma = 2$ [Fig. 2(d2)]. In this case, the GTAM operator is given by $\boldsymbol{J}^{(2)} = \boldsymbol{L} + 2\boldsymbol{S}$ [Fig.

2(c2)]. The corresponding pseudospin eigenstates of $S_z$ are $|\psi_2^\Gamma\rangle = (e^{i\pi 2/3}\ e^{i\pi 4/3}\ 1)^T$ and $|\psi_3^\Gamma\rangle = (e^{i\pi 4/3}\ e^{i\pi 2/3}\ 1)^T$. The commutation between the effective Hamiltonian $H_{\text{eff}}^\Gamma(\boldsymbol{p})$ and z-component $J_z^{(2)} = L_z + 2S_z$ is: $\left[H_{\text{eff}}^\Gamma, J_z^{(2)}\right] = 0$, indicating $J_z^{(2)}$ is also a conserved quantity for the QBTP. In contrast, the conventional TAM operator $J_z = L_z + S_z$ is not conserved in this case, as $\left[H_{\text{eff}}^\Gamma, J_z\right] = -2it(k_x^2 - k_y^2)S_y + 4itk_xk_yS_x \neq 0$. Thus, by introducing the winding number into the TAM, we show the GTAM conservation in lattices with band singularities beyond the linear dispersive Dirac point.

In a series of our previous work, we have shown that the conservation of the conventional TAM $\boldsymbol{J} = \boldsymbol{L} + \boldsymbol{S}$ in Dirac-like lattices with LBTPs, which leads to the topological charge conversion when one pseudospin state $s$ is initially excited with an OAM-carrying beam of topological charge $l$ [28–30,32]. Such a conversion obeys the TAM conservation expressed as $l + s = l' + s'$, where $(l, s)$, $(l', s')$ are OAM and PSAM at input and output, respectively. In the following, we shall demonstrate that such a TAM is not conserved for the QBTP in a Kagome lattice, but instead, the GTAM $\boldsymbol{J}^{(w)} = \boldsymbol{L} + w\boldsymbol{S}$ is conserved.

Our photonic Kagome lattice is established in a nonlinear crystal (Strontium-Barium Niobate (SBN:61)) via a multi-beam optical induction method [28,29]. A typical example obtained in experiment is shown in Fig.2(a2), with a lattice spacing $d = 12$ μm. Details of the lattice fabrication and experimental excitation schemes are provided in SI. To selectively excite pseudospin states $s = \pm 1/2$ for the LBTP, a donut-shaped triangular lattice beam carrying a global topological charge $l = \pm 1$ is employed [Fig. 3(a1)]. The probe beam comprises three vortex beams, aiming at three equivalent $K$- points in momentum space [Fig. 3(a2)]. By adjusting the phase differences of the three vortex beams through a spatial light modulator (SLM), the real-space phase of sublattices can be controlled to selectively excite the two pseudospin states $|\psi_1^K\rangle$ and $|\psi_2^K\rangle$ [Fig. 3(a3, a4)]. Since every $K$-points spectrum carries the same topological charge as the entire probe beam; thus, only one $K$-point is selected for interferometric phase measurement. The input interferogram reveals the desired topological charge ($l = 1$ and $l = -1$ for excitations $s = 1/2$ and $s = -1/2$, respectively) [Fig.3(b1, c1)]. As the designed probe beams excite only pseudospin modes around the LBTPs, the flat-band modes remain inactive and do not affect the propagation (see SI). After propagation in the photonic Kagome lattice ($z = 20$ mm), the output interferograms exhibit double vorticities with $l' = 2$ and $l' = -2$ [Fig. 3(b2, c2)]. These experimental results are consistent with the numerical simulations in Fig. 3(b3, c3). Additionally, to reveal the angular momentum conservation law, we numerically decompose the output beam into two pseudospin components.

When the pseudospin $s = 1/2$ ($s = -1/2$) is initially excited with a topological charge $l = 1$ ($l = -1$), the unexcited component $s' = -1/2$ ($s' = 1/2$) evolves into a vortex with $l' = 2$ ($l' = -2$) [Fig. 3(b5, c4)], while the initially excited one $s' = 1/2$ ($s' = -1/2$) remains unchanged [Fig. 3(b4, c5)]. These results confirm the relation $l + s = l' + s'$, thereby proving that $J_z^{(1)} = l + ws$ is a conserved quantity for the LBTP with winding number $w = 1$.

Next, we present the experimental results of GTAM conservation for the case of QBTP. The QBTP possesses pseudospin-1/2 and carries winding number $w = 2$, leading to topological charge conversion according to $l + 2s = l' + 2s'$. To excite the pseudospin states $s = \pm 1/2$ of the QBTP, the probe beam [Fig. 4(a1)] is constructed by interfering six vortex beams with the same topological charge $l = 1$ (or $l = -1$), whose wavevectors point towards all six equivalent $\Gamma_1$- points in the second BZ [Fig. 4(a2)]. The initial excitation $|\psi_2^\Gamma\rangle$ and $|\psi_3^\Gamma\rangle$, correspond to pseudospin $s = 1/2$ (or $s = -1/2$) [Fig. 4(a3, a4)], can be selectively excited by controlling the phase difference of the vortex beams. Since all six spectra possess identical topological charge, only the spectrum at one $\Gamma_1$-point is chosen for phase measurement. As shown in Figs. 4(b1, c1), the interferograms of the probe beam at input illustrate the initial excitation conditions similar to those in Fig. 3(b1, c1). In this configuration, the uppermost dispersive band remains unexcited (see SI). After 20mm propagation through the lattice, the output interferograms exhibit three vortices with the same vorticity as the input. These results agree well with numerical simulations [Figs. 4(b3, c3)], where a singly-charged probe beam also evolves into a triply-charged vortex. The slight deformation in the output patterns shown in the insets of Figs. 4(b2, b3) and 4(c2, c3) is attributed to the intrinsic instability of higher-order optical vortices [41].

To further illustrate GTAM conservation, we numerically decompose the output beam into two pseudospin components, as shown in Figs. 4(b4, b5) and 4(c4, c5). When the initial pseudospin $s = 1/2$ ($s = -1/2$) is excited with topological charge $l = 1$ ($l = -1$), the unexcited component $s' = -1/2$ ($s' = 1/2$) transforms into a vortex with $l' = 3$ ($l' = -3$), while the initially excited component retains its topological charge $l' = 1$ ($l' = -1$). Clearly, each component obeys the GTAM conservation rule: $l + ws = l' + ws'$, with $w = 2$ for the QBTP. Thus, the outcome from topological conversion for QBTP and LBTP is clearly different, even though both exhibit pseudospin-1/2, due to their distinct winding numbers at the singular BTPs. Together, our observations confirm that GTAM is indeed a conserved quantity in Kagome lattices hosting both linear and nonlinear BTPs.

The GTAM conservation can be applied to nonlinear BTPs with higher-pseudospin. For instance, for a three-band model with pseudospin-1 hosting parabolic intersections, the corresponding effective Hamiltonian can be written as

$$H_{\text{eff}}^{(2,3)}(\boldsymbol{p}) = \kappa^{(2,3)} \begin{pmatrix} 0 & (p_x - ip_y)^2 & 0 \\ (p_x + ip_y)^2 & 0 & (p_x - ip_y)^2 \\ 0 & (p_x + ip_y)^2 & 0 \end{pmatrix}, \quad (6)$$

where $\kappa^{(2,3)}$ is the lattice geometry-dependent coefficient. Equation (6) can be recast in a compact form as $H_{\text{eff}}^{(2,3)}(\boldsymbol{p}) = \sqrt{2}\kappa^{(2,3)} p^2 \left(S_x^{(3)}\cos(2\theta) + S_y^{(3)}\sin(2\theta)\right)$, with $\theta = \arctan(p_y/p_x)$, where $S_x^{(3)}$, and $S_y^{(3)}$ denote the $x, y$ components of the pseudospin-1 operator $\boldsymbol{S}^{(3)}$ By solving the eigenvalue of $H_{\text{eff}}^{(2,3)}(\boldsymbol{p})$, we obtain the band structure $E_0^{(2,3)} = 0$, and $E_\pm^{(2,3)} = \pm\sqrt{2}\kappa^{(2,3)} p^2$, where two quadratically dispersive bands touch with a flat band. (see details in SI). The GTAM operator and its $z$-component for this system are defined as $\boldsymbol{J}^{(2,3)} = \boldsymbol{L} + 2\boldsymbol{S}^{(3)}$ and $J_z^{(2,3)} = L_z + 2S_z^{(3)}$, respectively, with winding number $w = 2$. Remarkably, $J_z^{(2,3)}$ is also a conserved quantity for such a pseudospin-1 BTP with a quadratic dispersion, as it satisfies $\left[H_{\text{eff}}^{(2,3)}(\boldsymbol{p}), J_z^{(2,3)}\right] = 0$.

For the general case of multi-fold singular BTPs with higher-pseudospin and higher-winding numbers, we consider an arbitrary higher-order dispersion ($m$th-order) in an $n$-band model ($n > 2$). The Hamiltonian can be written as:

$$H_{\text{eff}}^{(m,n)}(\boldsymbol{p}) = \kappa^{(m,n)} p^m \left(S_x^{(n)}\cos(m\theta) + S_y^{(n)}\sin(m\theta)\right), \quad (7)$$

where $\kappa^{(m,n)}$ is the geometrical coefficient, the indices $(m, n)$ mark the band dispersion order and the band number. $S_x^{(n)}$ and $S_y^{(n)}$ are the $x$ and $y$ components of the PSAM operator $\boldsymbol{S}^{(n)}$, respectively. $\boldsymbol{S}^{(n)}$ acts on the number of $n$-fold touching bands, and varying $\boldsymbol{S}^{(n)}$ yields different Hamiltonians for the $n$-fold BTP. By applying Eq. (2), the pseudomagnetic-field winding number of the higher-pseudospin BTP is calculated as $w = m$. So, we can define the GTAM operator as $\boldsymbol{J}^{(m,n)} = \boldsymbol{L} + w\boldsymbol{S}^{(n)}$, which is a new conserved quantity of angular momentum characterizing the excitations around the BTPs. The commutation relation between $H_{\text{eff}}^{(m,n)}$ and $J_z^{(m,n)} = L_z + wS_z^{(n)}$ is well satisfied: $\left[H_{\text{eff}}^{(m,n)}, J_z^{(m,n)}\right] = 0$, indicating the $z$-component of the GTAM $J_z^{(m,n)}$ always remains conserved under the protection of rotational symmetry for such a generalized case.

In summary, we have established a generalized total angular momentum (GTAM) that remains conserved in discrete lattices hosting nonlinear BTPs, extending angular-momentum conservation beyond the Dirac paradigm. GTAM naturally incorporates the topological winding number of the singularity and reduces to the conventional TAM for linear Dirac points. Using a photonic Kagome lattice, we experimentally verified GTAM conservation through pseudospin–to-OAM conversion at both LBTPs and QBTPs. Because this framework applies broadly to multi-fold BTPs with arbitrary pseudospin textures and higher winding numbers, it reveals a general correspondence between angular momentum and topology in discrete systems. These results enrich the fundamental understanding of topological band singularities and open new avenues for manipulating angular momentum in photonic and condensed-matter platforms, including three-dimensional topological materials beyond Dirac and Weyl paradigm [26,39].

**Acknowledgments**

We thank Daniel Leykam for helpful discussion. This research is supported by the National Key R&D Program of China under Grant No. 2022YFA1404800, the National Natural Science Foundation of China (W2541003, 12134006, 12274242, 12374309, 12250410236, 12504449), and the QuantiXLie Center of Excellence, a project co-financed by the Croatian Government and European Union through the European Regional Development Fund - the Competitiveness and Cohesion Operational Programme (Grant KK.01.1.1.01.0004). D.B. acknowledges support from the 66 Postdoctoral Science Grant of China and the Ministry of Human Resources and Social Security of China (Grant WGXZ2023110). S.L. acknowledges support from the Postdoctoral Fellowship Program (Grade C) of China Postdoctoral Science Foundation (GZC20252211) and the China Postdoctoral Science Foundation - Tianjin Joint Support Program (2025T004TJ). R.M. acknowledges support from NSERC Discovery and the CRC program in Canada.

**Conflict of interests**

The authors declare no conflicts of interest and no competing financial interests.

**Contributions**

All authors participated in and contributed to this work. Z.C. and H.B. supervised the project.

**Reference**

[1] N. P. Armitage, E. J. Mele, and A. Vishwanath, Weyl and Dirac semimetals in three-dimensional solids, Rev. Mod. Phys. **90**, 015001 (2018).

[2] M. Z. Hasan, G. Chang, I. Belopolski, G. Bian, S.-Y. Xu, and J.-X. Yin, Weyl, Dirac and high-fold chiral fermions in topological quantum matter, Nat. Rev. Mater. **6**, 784 (2021).

[3] K. S. Novoselov, A. K. Geim, S. V. Morozov, D. Jiang, M. I. Katsnelson, I. V. Grigorieva, S. V. Dubonos, and A. A. Firsov, Two-dimensional gas of massless Dirac fermions in graphene, Nature **438**, 197 (2005).

[4] A. H. Castro Neto, F. Guinea, N. M. R. Peres, K. S. Novoselov, and A. K. Geim, The electronic properties of graphene, Rev. Mod. Phys. **81**, 109 (2009).

[5] M. Mecklenburg and B. C. Regan, Spin and the Honeycomb Lattice: Lessons from Graphene, Phys. Rev. Lett. **106**, 116803 (2011).

[6] M. I. Katsnelson, K. S. Novoselov, and A. K. Geim, Chiral tunnelling and the Klein paradox in graphene, Nat. Phys. **2**, 620 (2006).

[7] X. Jiang, C. Shi, Z. Li, S. Wang, Y. Wang, S. Yang, S. G. Louie, and X. Zhang, Direct observation of Klein tunneling in phononic crystals, Science **370**, 1447 (2020).

[8] Y. Zhang, Y.-W. Tan, H. L. Stormer, and P. Kim, Experimental observation of the quantum Hall effect and Berry's phase in graphene, Nature **438**, 201 (2005).

[9] K. Nomura, S. Ryu, M. Koshino, C. Mudry, and A. Furusaki, Quantum Hall Effect of Massless Dirac Fermions in a Vanishing Magnetic Field, Phys. Rev. Lett. **100**, 246806 (2008).

[10] M. Polini, F. Guinea, M. Lewenstein, H. C. Manoharan, and V. Pellegrini, Artificial honeycomb lattices for electrons, atoms and photons, Nat. Nanotechnol. **8**, 625 (2013).

[11] B. Bradlyn, J. Cano, Z. Wang, M. G. Vergniory, C. Felser, R. J. Cava, and B. A. Bernevig, Beyond Dirac and Weyl fermions: Unconventional quasiparticles in conventional crystals, Science **353**, aaf5037 (2016).

[12] K. Sun, H. Yao, E. Fradkin, and S. A. Kivelson, Topological Insulators and Nematic Phases from Spontaneous Symmetry Breaking in 2D Fermi Systems with a Quadratic Band Crossing, Phys. Rev. Lett. **103**, 046811 (2009).

[13] J.-X. Yin, B. Lian, and M. Z. Hasan, Topological kagome magnets and superconductors, Nature **612**, 647 (2022).

[14] M. Kang, S. Fang, L. Ye, H. C. Po, J. Denlinger, C. Jozwiak, A. Bostwick, E. Rotenberg, E. Kaxiras, J. G. Checkelsky, and R. Comin, Topological flat bands in frustrated kagome lattice CoSn, Nat. Commun. **11**, 4004 (2020).

[15] C. Wu, D. Bergman, L. Balents, and S. Das Sarma, Flat Bands and Wigner Crystallization in the Honeycomb Optical Lattice, Phys. Rev. Lett. **99**, 070401 (2007).

[16] C. D. Brown, S.-W. Chang, M. N. Schwarz, T.-H. Leung, V. Kozii, A. Avdoshkin, J. E. Moore, and D. Stamper-Kurn, Direct geometric probe of singularities in band structure, Science **377**, 1319 (2022).

[17] E. McCann, and V. I. Fal'ko, Landau-Level Degeneracy and Quantum Hall Effect in a Graphite Bilayer, Phys. Rev. Lett. **96**, 086805 (2006).

[18] S. Pujari, T. C. Lang, G. Murthy, and R. K. Kaul, Interaction-Induced Dirac Fermions from Quadratic Band Touching in Bilayer Graphene, Phys. Rev. Lett. **117**, 086404 (2016).

[19] K. Hejazi, C. Liu, and L. Balents, Landau levels in twisted bilayer graphene and semiclassical orbits, Phys. Rev. B **100**, 035115 (2019).

[20] F. De Juan, Non-Abelian gauge fields and quadratic band touching in molecular graphene, Phys. Rev. B **87**, 125419 (2013).

[21] R. Du, M.-H. Liu, J. Mohrmann, F. Wu, R. Krupke, H. Von Löhneysen, K. Richter, and R. Danneau, Tuning Anti-Klein to Klein Tunneling in Bilayer Graphene, Phys. Rev. Lett. **121**, 127706 (2018).

[22] J. Shah, and S. Mukerjee, Renormalization group study of systems with quadratic band touching, Phys. Rev. B **103**, 195118 (2021).

[23] J.-W. Rhim, K. Kim, and B.-J. Yang, Quantum distance and anomalous Landau levels of flat bands, Nature **584**, 59 (2020).

[24] M. Milićević, G. Montambaux, T. Ozawa, O. Jamadi, B. Real, I. Sagnes, A. Lemaître, L. Le Gratiet, A. Harouri, J. Bloch, and A. Amo Type-III and Tilted Dirac Cones Emerging from Flat Bands in Photonic Orbital Graphene, Phys. Rev. X **9**, 031010 (2019).

[25] Z. Gao, H. Zhao, T. Wu, X. Feng, Z. Zhang, X. Qiao, C.-K. Chiu, and L. Feng, Topological quadratic-node semimetal in a photonic microring lattice, Nat. Commun. **14**, 3206 (2023).

[26] C. Jörg, S. Vaidya, J. Noh, A. Cerjan, S. Augustine, G. von Freymann, and M. C. Rechtsman, Observation of Quadratic (Charge-2) Weyl Point Splitting in Near-Infrared Photonic Crystals, Laser Photonics Rev. **16**, 2100452 (2022).

[27] S. Vaidya, J. Noh, A. Cerjan, C. Jörg, G. Von Freymann, and M. C. Rechtsman, Observation of a Charge-2 Photonic Weyl Point in the Infrared, Phys. Rev. Lett. **125**, 253902 (2020).

[28] D. Song, V. Paltoglou, S. Liu, Y. Zhu, D. Gallardo, L. Tang, J. Xu, M. Ablowitz, N. K. Efremidis, and Z. Chen, Unveiling pseudospin and angular momentum in photonic graphene, Nat. Commun. **6**, 6272 (2015).

[29] X. Liu, S. Xia, E. Jajtić, D. Song, D. Li, L. Tang, D. Leykam, J. Xu, H. Buljan, and Z. Chen, Universal momentum-to-real-space mapping of topological singularities, Nat. Commun. **11**, 1586 (2020).

[30] S. Lei, S. Xia, D. Song, J. Xu, H. Buljan, and Z. Chen, Optical vortex ladder via Sisyphus pumping of Pseudospin, Nat. Commun. **15**, 7693 (2024).

[31] Z. Zhang, P. Kokhanchik, Z. Liu, Y. Shen, F. Liu, M. Liu, Y. Zhang, M. Xiao, G. Malpuech, and D. Solnyshkov, Spin-to-Orbital Angular Momentum Conversion in Non-Hermitian Photonic Graphene, arXiv:2504.03252.

[32] S. Lei, S. Xia, J. Wang, X. Liu, L. Tang, D. Song, J. Xu, H. Buljan, and Z. Chen, Mapping and Manipulation of Topological Singularities: From Photonic Graphene to T-Graphene, ACS Photonics **10**, 2390 (2023).

[33] P. Menz, H. Hanafi, D. Leykam, J. Imbrock, and C. Denz, Pseudospin-2 in photonic chiral borophene, Photon. Res. **11**, 869 (2023).

[34] D. Leykam and A. S. Desyatnikov, Conical intersections for light and matter waves, Adv. Phys. X **1**, 101 (2016).

[35] F. Diebel, D. Leykam, S. Kroesen, C. Denz, and A. S. Desyatnikov, Conical Diffraction and Composite Lieb Bosons in Photonic Lattices, Phys. Rev. Lett. **116**, 183902 (2016).

[36] C. Li and A. Miroshnichenko, Extended SSH Model: Non-Local Couplings and Non-Monotonous Edge States, Physics **1**, 2 (2018).

[37] X. Wan, S. Sarkar, S.-Z. Lin, and K. Sun, Topological Exact Flat Bands in Two-Dimensional Materials under Periodic Strain, Phys. Rev. Lett. **130**, 216401 (2023).

[38] I. V. Basistiy, V. Yu. Bazhenov, M. S. Soskin, and M. V. Vasnetsov, Optics of light beams with screw dislocations, Opt. Commun. **103**, 422 (1993).

[39] H. He, C. Qiu, X. Cai, M. Xiao, M. Ke, F. Zhang, and Z. Liu, Observation of quadratic Weyl points and double-helicoid arcs, Nat. Commun. **11**, 1820 (2020).

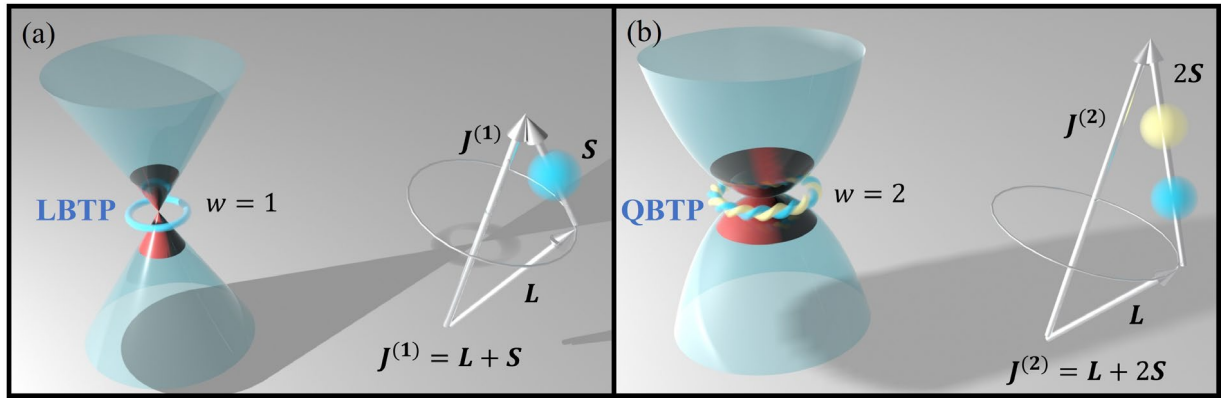

**Fig. 1 Schematic illustration of LBTPs and QBTPs and their corresponding GTAM (a)** Left: A LBTP with winding number $w = 1$; Right: Corresponding GTAM $J^{(1)} = L + S$. White arrows denote GTAM $J^{(1)}$, OAM $L$, and PSAM $S$, the blue sphere indicates the unit $S$ contribution. **(b)** Left: A QBTP with winding number $w = 2$; Right: Corresponding GTAM operator $J^{(2)} = L + 2S$. GTAM $J^{(2)}$, OAM $L$, and PSAM $S$ components are denoted by three white arrows. The blue and yellow spheres represent the doubled $S$ contribution due to winding number $w = 2$ of the QBTP.

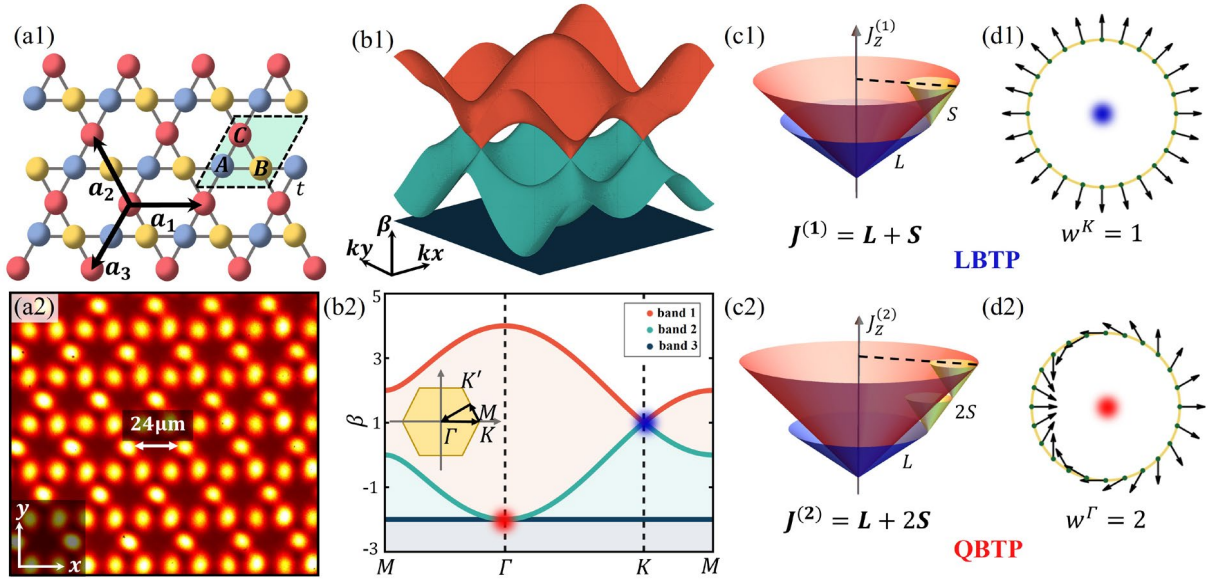

**Fig. 2 Theoretical analysis of the band structure and topological properties of LBTPs and QBTPs in a Kagome lattice.** **(a1)** Schematic of a Kagome lattice with one unit cell (enclosed by a dashed parallelogram) composed of three sublattices (A, B, and C), and primitive lattice vectors $\boldsymbol{a1}$, $\boldsymbol{a2}$, and $\boldsymbol{a3}$. **(a2)** Photonic Kagome lattice from experiment. **(b1)** Tight-binding band structure. **(b2)** Projected band structure along high-symmetry points indicated by black arrows in inset. Blue and red dots highlight the position of the LBTP and QBTP at $K$- and $\varGamma$- points, respectively. **(c1, c2)** Diagrams showing the conservation of GTAM for (c1) LBTPs and (c2) QBTPs. The blue, yellow, and red cones represent the OAM, PSAM, and GTAM, respectively. **(d1, d2)** Pseudomagnetic field winding number $w$ of BTPs: (d1) $w = 1$ for LBTP and (d2) $w = 2$ for QBTP. Yellow curves indicate the winding path of pseudomagnetic-field $\boldsymbol{B(p)}$.

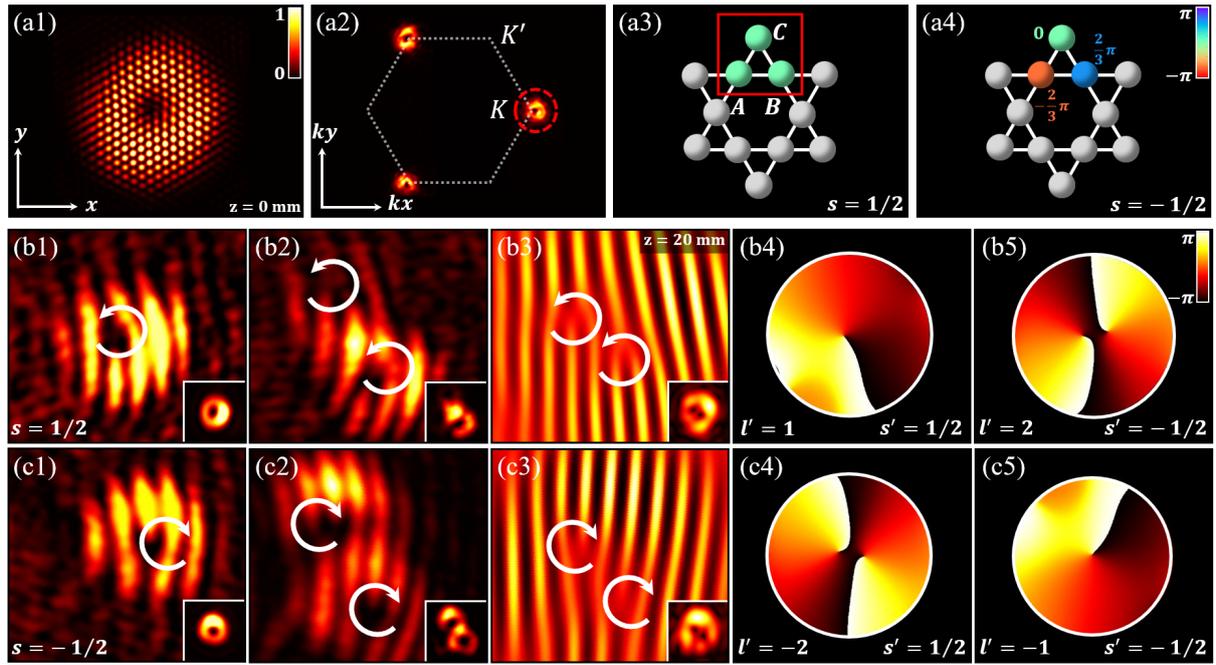

**Fig. 3 Experimental observation of pseudospin-to-OAM conversion at LBTPs. (a1, a2)** Input intensity distribution (a1) and corresponding spectrum (a2) of the vortex probe beam, exciting the pseudospin state $s$ of LBTP with initial topological charge $l$. **(a3, a4)** Phase distributions of pseudospin states $s = 1/2$ (a3), and $s = -1/2$ (a4) among three sublattice sites in one unit cell. **(b, c)** Experiment and numerical simulation for selective excitation of $s = 1/2$ and $s = -1/2$, with vortex probe beams of topological charge $l = 1$ and $l = -1$, respectively. **(b1, c1)** Input Interferogram from one spectral component at input, selected from one $K$-point of the probe beam as indicated by the red dashed circle in (a2). **(b2-b3, c2-c3)** Measured and numerically simulated output interferograms, highlighting that the topological charge of the probe beam increases to $l' = 2$ (b2, b3) and decreases to $l' = -2$ (c2, c3). **(b4-b5, c4-c5)** Corresponding numerically decomposed phase profiles of the output pseudospin components (b4, c4) $s' = 1/2$ and (b5, c5) $s' = -1/2$. Insets in (b1-b3, c1-c3) show the intensity distribution of the filtered spectrum component. The white dashed hexagon in (a2) marks the first BZ while white arrows in (b1-b3, c1-c3) indicate the position and helicity of the vortices.

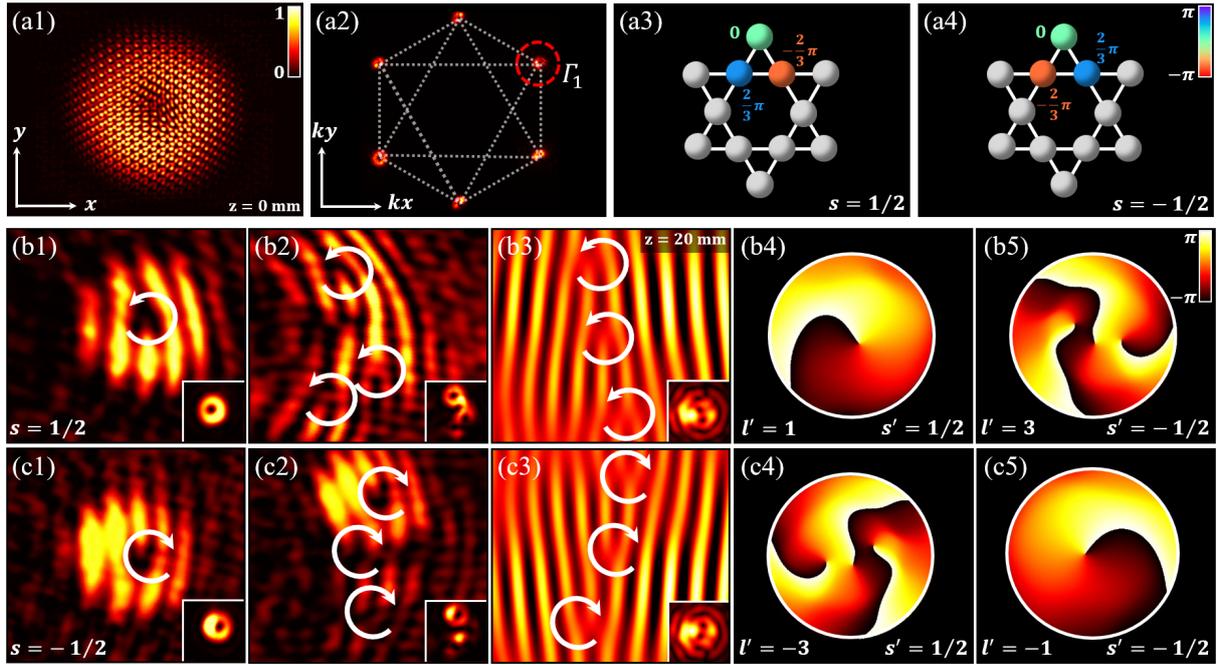

**Fig. 4 Experimental observation of pseudospin-to-OAM conversion at QBTPs. (a1, a2)** Input intensity distribution (a1) and corresponding spectrum (a2) of the vortex probe beam, exciting the pseudospin state $s$ of QBTP with initial topological charge $l$. **(a3, a4)** Phase distributions of pseudospin states $s = 1/2$ (a3), and $s = -1/2$ (a4) among three sublattice sites in one unit cell. **(b, c)** Experiment and numerical simulation for selective excitation of $s = 1/2$ and $s = -1/2$, with vortex probe beams of topological charge $l = 1$ and $l = -1$, respectively. **(b1, c1)** Input Interferogram from one spectral component at input, selected from one $\varGamma$-point in the second BZ of the probe beam as indicated by the red dashed circle in (a2). **(b2-b3, c2-c3)** Measured and numerically simulated output interferograms, highlighting that the topological charge of the probe beam increases to $l' = 3$ (b2, b3) and decreases to $l' = -3$ (c2, c3). **(b4-b5, c4-c5)** Corresponding numerically decomposed phase profiles of the output pseudospin components (b4, c4) $s' = 1/2$ and (b5, c5) $s' = -1/2$. Insets in (b1-b3, c1-c3) show the intensity distribution of the filtered spectrum component. The white dashed hexagon and triangles in (a2) marks the first and second BZs while white arrows in (b1-b3, c1-c3) indicate the position and helicity of the vortices.